\documentclass[]{article}
\begin{document}
\title{A tube concept in rubber viscoelasticity}

\author{Aleksey D. Drozdov\\
Lehrstuchl f\"{u}r Polymerwerkstoffe\\
Institut f\"{u}r Werkstoffwissenschaften\\
Universit\"{a}t Erlangen--N\"{u}rnberg,
Martensstrasse 7\\
D--91058, Erlangen, Germany}

\date{}
\maketitle

\begin{abstract}
A constitutive model is derived for the time-dependent response
of particle-reinforced elastomers at finite strains.
An amorphous rubbery polymer is treated as a network of long
chains linked by permanent junctions (chemical crosslinks,
entanglements and filler particles).
A strand between two neighboring junctions is thought of as a sequence of mers 
whose motion is restricted to some tube by surrounding macromolecules.
Unlike the conventional approach that presumes the cross-section
of the tube to be constant, we postulate that its radius 
strongly depends on the longitudinal coordinate.
This implies that a strand may be modeled as a sequence of segments
whose thermal motion is totally frozen by the environment (bottle-neck points
of the tube) bridged by threads of mers which go through all possible configurations during 
the characteristic time of a test.
Thermal fluctuations affect the tube's radius, which results
in freezing and activation of regions of high molecular mobility (RHMs).
The viscoelastic response of an elastomer is associated with
thermally activated changes in the number of RHMs in strands.
Stress--strain relations for a rubbery polymer at finite strains 
and kinetic equations for the concentrations of RHMs are 
developed by using the laws of thermodynamics.
At small strains these relations are reduced to the conventional 
integral constitutive equation in linear viscoelasticity
with a novel scaling law for relaxation times.
The governing equation is determined by 5 adjustable parameters
which are found by fitting experimental data in tensile dynamic tests
on a carbon black filled natural rubber vulcanizate.
\end{abstract}
\vspace*{5 mm}
\noindent
{\bf Key-words:} Particle-reinforced elastomers, Viscoelasticity, Tube concept,
Thermal fluctuations, Dynamic tests

\section{Introduction}

This study is concerned with the viscoelastic behavior of unfilled and
particle-reinforced rubbers in isothermal tests.
The time-dependent response of elastomers and rubbery polymers has attracted
essential attantion in the past decade, which may be explained by numerous 
applications of rubber-like materials in industry
(vehicle tires, shock absorbers, earthquake bearings, seals, flexible joints, 
solid propellants, etc.).
Despite noticeable progress in this field
(see, e.g.,
Govenjee and Simo, 1992;
Ozupek and Becker, 1992;
Gonzonas, 1993;
Johnson and Stacer, 1993;
Le Tallec et al., 1993;
Zdunek, 1993;
Hausler and Sayir, 1995;
Johnson et al., 1995;
Aksel and H\"{u}bner, 1996;
Holzapfel and Simo, 1996;
Lion, 1996, 1997, 1998;
Spathis, 1997;
Bergstr\"{o}m and Boyce, 1998;
Ha and Schapery, 1998;
Reese and Govinjee, 1998;
Septanika and Ernst, 1998a,b;
Ernst and Septanika, 1999;
Mieche and Keck, 2000;
Wu and Liechti, 2000;
Haupt and Sedlan, 2001),
some important questions remain, however, obscure.

Numerous observations in uniaxial and biaxial tests evidence that commercial rubbery
vulcanizates reinforced with carbon black and silica, as well as solid propellants 
filled with micrometer-size hard particles demonstrate a pronounced
time-dependent response which may be associate with their viscoelastic properties.
For examples, we refer to recent studies by Hausler and Sayir (1995),
Lion (1996--1998), Bergstr\"{o}m and Boyce (1998),
Septanika and Ernst (1998b), Mieche and Keck (2000),
Wu and Liechti (2000), and Haupt and Sedlan (2001).
An unresolved problem remains about the mechanism for stress relaxation
at the molecular lovel.

To answer this question, Aksel and H\"{u}bner (1996) carried our tensile relaxation
tests on crosslinked polybutadiene rubber reinforced with glass beads 
at ambient temperature with various volume fractions of filler (from 0 to 45 \%).
Obvervations evidenced that there was no relaxation in the unfilled rubber, 
and the level of relaxed stress increased with the content of reinforcement.
They suggested that slippage of polymeric chains along the filler surface may serve 
as a mechanism for stress relaxation in the linear regime,
whereas detachment of chains from the surfaces of beads and creation of valuoles
characterized the nonlinear viscoelasticity of rubber.

Liang et al. (1999) reported experimental data in dynamic tensile tests 
(with frequency 1 Hz) on ternary composite of polypropylene and 
EPDM (ethylene propylene diene monomers)  elastomer filled with glass beads. 
They demonstrated that (i) treatment of beads by silane coupling agent did not 
affect the mechanical damping of specimens (measured in terms of the loss modulus)
and (ii) the loss modulus of the compound with untreated beads decreased with 
the concentration of filler [in contrast with the observations by Aksel and H\"{u}bner (1996)].

These conclusions were recently questioned by Leblanc and Cartault (2001), who measured
dynamical shear response of uncured styrene--butadiene rubber reinforced with 
various amounts of carbon black and silica and found that the compounds exhibited 
practically no viscoelasticity or rather limited time-dependent response.
This implies that crosslinks play the key role in viscoelasticity of rubber,
while the effect of filler particles is relatively weak.

Clarke et al. (2000) carried out tensile relaxation tests on two 
unfilled elastomers (natural polyisoprene rubber
with polysulphur crosslinks and synthetic polyacrylate rubber with carbon--carbon
and carbon--oxygen crosslinks) at the temperature $T=80$ $^{\circ}$C.
At a relatively small strain ($\epsilon=0.05$)
the natural rubber demonstrated significant mechanical relaxation
(which is observed during at least 10$^{6}$ s), whereas the synthetic elastomer
revealed no relaxation.
Clarke et al. (2000) associated slow relaxation of natural rubber with isomerisation of
polysulphur crosslinks, but could not explain stress relaxation in natural rubber
observed in short-term mechanical tests, as well as its absence in polyacrylate
elastomer.

Two approaches may be mentioned to the description of the time-dependent response in
rubbery polymers.
The first was proposed about half a century ago by Green and Tobolsky (1946)
and, afterwards, revised by Yamamoto (1956), Lodge (1968) and Tanaka and Edwards (1992),
to mention a few, and applied to describe the viscoelastic behavior of rubbery polymers 
by Drozdov (1997) and Ernst and Septanika (1999).
According to the concept of transient networks, a rubbery polymer 
is modeled as a network of long chains bridged by temporary junctions
(which are associated with chemical and physical crosslinks and entanglements).
A strand whose ends are linked to neighboring junctions is treated as 
an active one.
Snapping of an end of a strand from a junction (breakage of a crosslink
or disentanglement of a chain) is thought of as its breakage (transition from 
the active state of the strand to its dangling state).
When a dangling chain captures a junction (which reflects creation of a new
physical crosslink or entanglement), a new active strand merges with the network.
Breakage and reformation of active strands are treated as thermally
activated processes: attachment and detachment events occur at random times 
as they are agitated by thermal fluctuations.
A disadvantage of this model is that it does not explain observations by
Leblanc and Cartault (2001): the theory of transient networks implies an increase
in the strength of relaxation in reinforced rubbery compounds independently of the
degree of crosslinking, because particles of filler play the role of extra junctions
in the network.
It is worth mentioning two shortcomings of this theory from the theoretical standpoint:
(i) it associates viscoelasticity with the micro-motion of strands, which contradicts
the conventional opinion that relaxation of stresses reflects {\it local} segmental 
motion in the chain backbone (Sperling, 1996),
and (ii) this approach is grounded on the hypothesis that the stress-free state of a reformed
strand coincides with the deformed state of the network at the instant of their merging.
The latter is tantamount to implicit ascribing some elastic features to long chains
[see a recent discussion of this issue for microemulsions with telechelic polymers
in Filali et al. (2001)], whereas the conventional theory of rubber elasticity disregards 
the contribution of mechanical energy into the free energy of chains compared 
to that of configurational entropy (Treloar, 1975).

The other concept was recently suggested by Clarke et al. (2000) who associated the viscoelastic
behavior of elastomers with mechanically induced rearrangement (isomerisation) of crosslinks.
The shortcomings of this approach are that (i) it contradicts experimental data by 
Aksel and H\"{u}bner (1996) (because it disregards the effect of filler)
and (ii) this model does not explain the presence of different relaxation modes
observed in a number of mechanical tests (because crosslinks are identical, their 
rearrangement results in a unique relaxation time inversely proportional to the characteristic
time of rearrangement), and (iii) it fails to describe the viscoelastic behavior of
uncrosslinked polymers (uncured rubbers or thermoplastics above the glass transition temperature).

The objective of this study is three-fold:
(i) to develop a micromechanical model which does not contradict 
the above experimental data,
(ii) to find adjustable parameters in constitutive equations by fitting observations
in dynamic tests with a wide range of frequencies (3 decades) at various temperatures
(from $-20$ to 100 $^{\circ}$C),
and (iii) to demonstrate that the material constants are altered with temperature 
in a physically plausible way.
By no means this model is treated as an alternative for existing theories, but
rather as an accompanying concept.
This is a reason why we focus on observations in short-term tests (the frequency, $f$, 
changes from $10^{-1}$ to $10^{2}$ Hz) with relatively small strains (less than 10 \%),
by assuming that (i) the theory of transient networks with mechanically induced rates for
breakage and reformation of active strands can correctly predict experimental data 
in the region of larger strains, where disentanglement of chains and dewetting of the host polymer 
from the filler surfaces become essential, 
and (ii) the theory of junctions' rearrangement can be adequately applied
to the analysis of the time-dependent response in long-term tests (with the characteristic
time of the order of $10^{6}$ s).

At first glance, approximation of experimental data in isothermal dynamic tests 
with small strains is an oversimplified task.
Assuming the stress, $\sigma(t)$, at an arbitrary time $t\geq 0$ to depend linearly
on the strain history, $\epsilon(\tau)$, $\tau\in [0,t]$, and referring to
the Risz theorem for the presentation of a linear functional, one can write
\begin{equation}
\sigma(t)=E\Bigl [\epsilon(t)-\int_{0}^{t} R(t-\tau)\epsilon(\tau) d\tau\Bigr ],
\end{equation}
where $E$ is an elastic modulus and $R(t)$ is a relaxation kernel.
Expanding the function $R(t)$ into a truncated Prony series,
\begin{equation}
R(t)=\sum_{N=1}^{N_{1}} \xi_{N}\exp \biggl (-\frac{t}{\tau_{N}}\biggr ),
\end{equation}
one finds that given a number of terms, $N_{1}$, and a series of relaxation times,
$\tau_{N}$, the parameters $E$ and $\xi_{N}$ may be uniquely determined by matching
the storage and loss moduli with the help of the least-squares algorithm.
This approach perfectly works when we aim to {\it approximate} observations 
at some temperature and amplitude of oscillations.
In a more complicated case which is addressed here, when observations are available
at several temperatures and amplitudes of strains (the response of particle-reinforced
rubber vulcanizates is extremely sensitive to the amplitude of oscillations even
for strains of the order of several per cent) and the objective is to {\it evaluate} the
effect of thermomechanical factors on material parameters, 
phenomenological equations (1) and (2) totally  fail, because they provide no 
information about changes in the relaxation times, $\tau_{N}$, with temperature and strain.

To shed some light on the effects of temperature and strain on the relaxation spectrum, 
we develop a model for the time-dependent response of elastomers at finite strains
(i) that is reduced to Eq. (1) in the case of small strains, and (ii) which
expresses the quantities $\xi_{N}$ and $\tau_{N}$ in Eq. (2) in terms of some
micromechanical parameters, whose dependence on thermomechanical factors
appears to be physically plausible.
This model contains only five material constants which are determined by fitting observations.

According to the statistical theory of macromolecules,
a rubbery polymer is thought of as a network of long chains linked by permanent
junctions (crosslinks, entanglements and van der Waals forces).
A strand between two neighboring junctions is thought of as a set
of large number of mers connected in sequel.
The classical theory of rubber elasticity treats an individual strand as an
object independent of its surrounding, which implies that the configurational 
entropy (associated with the number of configurations available for statistically 
independent segments in the strand) provides the main contribution into the free 
energy per strand.
It was recognized four decades ago that micro-motion of segments is confined
to the neighborhood of some ``mean configuration" by surrounding macromolecules
which suppress their thermal fluctuations.
This confinement is traditionally accounted for in the framework of the tube 
concept which presumes that a chain can freely move only 
within some tube created by neighboring macromolecules (Doi and Edwards, 1986).
For a detailed discussion of constrained fluctuations, the reader is referred to Everaers (1998).

The tube is conventionally treated as a curvilinear cylinder with a constant
cross-sectional area.
The main hypothesis of the present study is that small-scale local heterogeneities
dramatically affect micro-motion of a strand, which implies that the
radius of the cylinder strongly depends on the longitudinal coordinate.
This means that the surface of the cylinder is extremely wavy, and the tube may be treated 
as a sequence of regions with high mobility of segments bridged by bottle-neck
regions, where the molecular mobility is strongly suppressed.
The sketch depicted in Figure 1 demonstrates the position of the tube
and the location of regions of high mobility (RHMs) linked by regions
of suppressed mobility at the current time $t$.
Thermal fluctuations of surrounding macromolecules change the positions 
of RHMs noticeably even after a small interval of time, $\Delta t$,
which implies that the current properties of a strand may be described 
in terms of the statistical distribution of RHMs only.

To make the model tractable, we introduce the distribution, $p_{N}$, of strands 
with various numbers of RHMs, $N$, in an equilibrated sample 
(in the stress-free state at time $t=0$).
Application of external loads activates the network, which implies that
the concentrations of RHMs in strands are altered.
At an arbitrary instant $t\geq 0$, the state of a strand (which initially has $N$ RHMs)
is described by the current number of regions of high mobility, $N_{\rm a}(t)$.
Suppression of RHMs by surrounding molecules, as well as activation of
frozen segments (agitated by thermal fluctuations occuring at random times)
are assumed to serve as a basic mechanism for the viscoelastic response 
of an elastomer at the micro-level.

The study is organized as follows.
Deformation of a strand is described in Section 2.
The mechanical energy of a network is determined in Section 3.
Stress--strain relations at finite strains and a kinetic equation for the
function $N_{\rm a}(t)$ are derived in Section 4 by using the laws of thermodynamics.
The constitutive equations are simplified for the case of small strains in Section 5.
Uniaxial tension of a specimen is analyzed in Section 6.
Results of numerical simulation are compared with observations in dynamic tests
on a carbon black filled natural rubber vulcanizate in Section 7.
The experiments exposed in the introductory section are shortly discussed
in terms of the model in Section 8.
Some concluding remarks are formulated in Section 9.

\section{Deformation of a strand}

We consider a strand linking two junctions, $A_{1}$ and $A_{2}$,
which contains $N$ regions of high molecular mobility in the stress-free state 
at the initial instant $t=0$ and $N_{\rm a}(t)$ RHMs at time $t\geq 0$.
Introducing the dimensionless increment of the number of RHMs,
\[ n(t)=\frac{N-N_{\rm a}(t)}{N}, \]
we find that
\begin{equation}
N_{\rm a}(t)=N [1-n(t)].
\end{equation}
It is worth noting that the function $n(t)$ may accept both positive ($N_{\rm a}(t)<N$)
and negative ($N_{\rm a}(t)<N$) values.

Let $e$ be a strain from the stress-free state of a strand to its deformed state.
Assuming that suppressed regions are not deformed, we find that the strain $e$ 
equals the sum of strains for active RHMs,
\[ e=N_{\rm a}e_{\rm a}. \]
This equality together with Eq. (3) implies that
\begin{equation}
e_{\rm a}=\frac{e}{N(1-n)}.
\end{equation}
The mechanical energy of suppressed regions (totally frozen segments of a strand) 
is assumed to vanish.
Active RHMs are modeled as linear elastic solids with the strain energy
\begin{equation}
\frac{1}{2} \mu e_{\rm a}^{2}, 
\end{equation}
where the rigidity per bond $\mu$ has the dimension of energy.
Two remarks are necessary regarding Eq. (5).
First, this relation presumes small deformations of RHMs.
This hypothesis was confirmed by recent molecular dynamic simulations
(Bergstr\"{o}m and Boyce, 2001) which demonstrated that axial elongation of
a specimen with macro-strains up to 100 \% implied micro-strains per node which did
not exceed 5 \% (for chains containing 200 atoms, which is in quantitative 
agreement with data obtained by fitting observations in our study, see Section 7).
Second, at small strains, the contributions of mechanical
and statistical elasticity into the free energy of a polymer cannot be distinguished
(Ferry, 1980), which implies that $\mu=\mu(T)$ is the rigidity associated with both
(entropic or mechanical) components of the potential energy of deformation.

The strain energy of a strand, $w_{N}$, is determined as the sum of the strain
energies for individual RHMs.
It follows from Eqs. (4) and (5) that
\begin{equation}
w_{N}(e) =\frac{1}{2}\mu N_{\rm a} e_{\rm a}^{2}=\frac{\mu e^{2}}{2 N(1-n)}.
\end{equation}
To express the strain in a strand, $e$, in terms of
the Cauchy deformation tensor for the network, $\hat{C}$,
we suppose that in the stress-free state
the strand has a small end-to-end length $\delta_{N}$
and it is directed along the unit guiding vector $\bar{l}$,
\[ \bar{r}_{2}^{\circ}-\bar{r}_{1}^{\circ}=\delta_{N} \bar{l}, \]
where $\bar{r}_{1}^{\circ}$ and $\bar{r}_{2}^{\circ}$ 
are initial radius vectors of the end points, $A_{1}$ and $A_{2}$.
At some instant $t\geq 0$, the junctions occupy points with the radius vectors
\[ \bar{r}_{1}(t)=\bar{r}_{1}^{\circ}+\bar{u}(t,\bar{r}_{1}^{\circ}),
\qquad
\bar{r}_{2}(t)=\bar{r}_{2}^{\circ}+\bar{u}(t,\bar{r}_{2}^{\circ}), \]
where $\bar{u}(t,\bar{r})$ is the displacement vector
at point $\bar{r}$ for transition from the stress-free state
to the deformed state at time $t$.
The current position of the strand is determined by the end-to-end vector
\[ \bar{R}(t) = \bar{r}_{2}(t)-\bar{r}_{1}(t)
= \delta_{N}\bar{l}+[\bar{u}(t,\bar{r}_{1}^{\circ}+\delta_{N}\bar{l})
-\bar{u}(t,\bar{r}_{1}^{\circ})]. \]
Neglecting terms beyond the first order of smallness compared
to $\delta_{N}$, we find that
\[ \bar{R}(t)=\delta_{N}\bar{l} \cdot [\hat{I}
+\bar{\nabla}^{\circ}\bar{u}(t,\bar{r}_{1}^{\circ})], \]
where $\bar{\nabla}^{\circ}$ is the gradient operator
in the stress-free state,
$\hat{I}$ is the unit tensor and the dot stands for inner product.
In terms of the radius vector in the actual state, $\bar{r}$,
this equality reads
\begin{equation}
\bar{R}(t)=\delta_{N} \bar{l}\cdot \bar{\nabla}^{\circ}\bar{r}(t)
=\delta_{N}[ \bar{\nabla}^{\circ}\bar{r}(t) ]^{\top} \cdot \bar{l},
\end{equation}
where $\top$ stands for transpose.
The end-to-end length of the strand, $ds$, is given by
\[ ds^{2}=\bar{R}\cdot \bar{R}. \]
It follows from this equality and Eq. (7) that
\begin{equation}
ds^{2}(t,\bar{l}) = \delta_{N}^{2} \bar{l}\cdot \hat{C}(t)\cdot \bar{l},
\end{equation}
where
\begin{equation}
\hat{C}(t)=\bar{\nabla}^{\circ}\bar{r}(t)
\cdot [ \bar{\nabla}^{\circ}\bar{r}(t) ]^{\top}
\end{equation}
is the Cauchy deformation tensor for transition from
the stress-free state of the network to its deformed state at time $t$.

The extension ratio for a strand, $\lambda$, equals the ratio
of the current end-to-end length, $ds$, to that of the strand
in an ``activated'' stress-free state, $ds^{\circ}$.
The latter state is a state of the strand which is suddenly
unloaded, but for which the number of RHMs coincides with its current 
value, $N_{\rm a}(t)$.
It differs from the equilibrium stress-free state of the strand,
where the number of RHMs equals $N$.
The difference between the end-to-end lengths of the strand
in the activated state, $ds^{\circ}(t)$, and in the equilibrium state,
$ds(0)=\delta_{N}$, determines the end-to-end elongation
driven by changes in the number of RHMs.
We assume that the transformation-induced end-to-end elongation of a chain
is proportional to the number of frozen RHMs, $N-N_{\rm a}(t)$, which implies that
\begin{equation}
ds^{\circ}(t,\bar{l})=\delta_{N}+N n(t,\bar{l})\delta^{\circ},
\end{equation}
where $\delta^{\circ}$ is an increment of the end-to-end length driven 
by transition of an individual RHM.
It follows from Eqs. (8) and (10) that
\begin{equation}
\lambda(t,\bar{l})=\frac{ds(t,\bar{l})}{ds^{\circ}(t,\bar{l})}
=\frac{[ \bar{l}\cdot\hat{C}(t)\cdot\bar{l}]^{\frac{1}{2}}}{1
+\eta_{N} n(t,\bar{l})}, 
\end{equation}
where
\[ \eta_{N}=\frac{N \delta^{\circ}}{\delta_{N}}. \]
Adopting the conventional hypothesis in rubber elasticity (which is tantamount
to the assumption about statistical independence of RHMs, see Treloar, 1975) that 
\[ \delta_{N}=\tilde{\delta}\sqrt{N}, \]
where the coefficient $\tilde{\delta}$ is independent of $N$,
we arrive at the formula
\begin{equation}
\eta_{N}=\tilde{\eta}\sqrt{N},\qquad
\tilde{\eta}=\frac{\delta^{\circ}}{\tilde{\delta}}.
\end{equation}
The strain in a strand is determined by the Hencky formula 
\[ e=\ln \lambda. \]
Substitution of expression (11) into this relation implies that
\begin{equation}
e(t,\bar{l}) = \frac{1}{2} \ln
\Bigl [ \bar{l}\cdot\hat{C}(t)\cdot\bar{l}\Bigr ]
-\ln \Bigl [ 1+\eta_{N} n(t,\bar{l})\Bigr ].
\end{equation}
Equation (13) establishes correspondence between the micro-deformation of a strand which is
described by the strain, $e$, and the macro-deformation of a network which is determined
by the Cauchy deformation tensor, $\hat{C}$.
The parameter $\eta_{N}$ in Eq. (13) may serve as a measure of coiling for strands:
for strongly coiled chains, transformation-induced elongation of a strand 
is negligible compared to its length and $\eta_{N}$ is close to zero, 
whereas for uncoiled chains $\eta_{N}$ has the order of unity.

\section{Strain energy density of a network}

In the stress-free state a rubbery polymer is modeled as a network of strands 
with various numbers of RHMs ($N=1,2,\ldots,N_{1}$), where $N_{1}$ is the maximal number
of regions of high mobility per strand.
Denote by $\Xi_{N}$ the concentration (per unit mass) of strands with $N$ RHMs (in thermal equilibrium 
in the stress-free state) and by 
\[ \Xi=\sum_{N=1}^{N_{1}} \Xi_{N} \]
the total number of strands per unit mass.
The distribution of strands is characterized by the function
\[ p_{N}=\frac{\Xi_{N}}{\Xi}. \]
Referring to the random energy model (see, e.g., Dyre, 1995),
we accept the quasi-Gaussian ansatz for the probability density $p_{N}$,
\begin{equation}
p_{N}=A\exp\Bigl [-\frac{(N-\tilde{N})^{2}}{2\Sigma^{2}}\Bigr ],
\end{equation}
where $\tilde{N}$ and $\Sigma$ are positive constants, and the parameter $A$
is determined by the condition
\[ \sum_{N=1}^{N_{1}} p_{N} =1.  \]
When $\tilde{N}\gg \Sigma$, we can identify $\tilde{N}$ with the average number
of RHMs in a strand and $\Sigma$ with the standard deviation of the number
of RHMs.

We accept the conventional assumption that the excluded-volume effect
and other multi-chain effects are screened for an individual chain
by surrounding macromolecules (Everaers, 1998).
This implies that the energy of interaction between strands
may be neglected (under the hypothesis of incompressibility)
and the strain energy of the network may be determined as the sum 
of the potential energies of deformation for individual strands.
Assuming the distribution of strands with various guiding vectors 
to be isotropic, we find the concentration (per unit mass) of strands with $N$ RHMs
and guiding vector $\bar{l}$,
\begin{equation}
X_{N}(\bar{l})=\frac{1}{4\pi} \Xi p_{N} \sin\vartheta d\vartheta d\varphi , 
\end{equation}
where $\vartheta$ and $\varphi$ are Euler's angles
that determine the position of the unit vector $\bar{l}$.
Multiplying the number of strands, $X_{N}(\bar{l})$, by their mechanical energy, $w_{N}$,
summing the results for various $N$ and $\bar{l}$,
and using Eqs. (6) and (15), we find the strain energy of the network
\begin{equation}
W(t) = \frac{\mu\Xi}{8\pi} \sum_{N=1}^{N_{1}} \frac{p_{N}}{N} \int_{0}^{2\pi} d\varphi
\int_{0}^{\pi} \frac{e^{2}(t,\vartheta,\varphi)}{1
-n(t,\vartheta,\varphi)} \sin\vartheta d\vartheta .
\end{equation}
Our purpose now is to find the derivative of the function $W$ with respect to time.
Because the parameters $\mu$, $\Xi$ and $p_{N}$ are, in general, 
functions of temperature, $T$, the derivatives of these quantities 
should be accounted for in the calculation of $dW/dt$ for an arbitrary 
thermo-mechanical process. 
To simplify the analysis, we confine ourselves to small changes in
temperature and presume that the concentrations of strands, $\Xi_{N}$,
with various numbers of RHMs, $N$, are affected by temperature rather weakly.
This implies that only the derivative of the average rigidity per bond, $\mu$, is
taken into account in the formula for the derivative of $W$.
Differentiation of Eq. (16) results in
\begin{equation}
\frac{dW}{dt}(t) = \frac{\mu\Xi }{4\pi}\sum_{N=1}^{N_{1}}\frac{p_{N}}{N}
\int_{0}^{2\pi} d\varphi \int_{0}^{\pi}
\frac{e(t,\vartheta,\varphi)}{1-n(t,\vartheta,\varphi)}
\frac{\partial e}{\partial t}(t,\vartheta,\varphi)
\sin\vartheta d\vartheta
+J_{0}(t)\frac{dT}{dt}(t)+J_{1}(t),
\end{equation}
where
\begin{eqnarray}
J_{0}(t) &=& \frac{\Xi}{8\pi} \frac{d\mu}{dT}
\sum_{N=1}^{N_{1}} \frac{p_{N}}{N} \int_{0}^{2\pi} d\varphi
\int_{0}^{\pi} \frac{e^{2}(t,\vartheta,\varphi)}{1
-n(t,\vartheta,\varphi)} \sin\vartheta d\vartheta ,
\nonumber\\
J_{1}(t) &=& \frac{\mu\Xi }{8\pi}\sum_{N=1}^{N_{1}}\frac{p_{N}}{N}
 \int_{0}^{2\pi} d\varphi \int_{0}^{\pi}
\frac{e^{2}(t,\vartheta,\varphi)}{[1-n(t,\vartheta,\varphi)]^{2}}
\frac{\partial n}{\partial t}(t,\vartheta,\varphi)
\sin\vartheta d\vartheta. 
\end{eqnarray}
The function $J_{0}(t)$ in Eq. (18) determines the increment of the strain energy of 
a network associated with the dependence of the average rigidity per RHM, $\mu$,
on temperature, whereas the function $J_{1}(t)$ characterizes the increment
of the potential energy of deformation driven by suppresion and thermal activation of RHMs.
Differentiation of Eq. (13) results in
\begin{equation}
\frac{\partial e}{\partial t}(t,\bar{l}) =
\frac{1}{2[\bar{l}\cdot\hat{C}(t)\cdot\bar{l}]}
\Bigl [\bar{l}\cdot\frac{d\hat{C}}{dt}(t)\cdot\bar{l}\Bigr ]
-\frac{\eta_{N}}{1+\eta_{N} n(t,\bar{l})}
\frac{\partial n}{\partial t}(t,\bar{l}).
\end{equation}
It follows from Eq. (9) that
\[ \frac{d\hat{C}}{dt}(t)=\bar{\nabla}^{\circ}\bar{v}(t)\cdot
[\bar{\nabla}^{\circ}\bar{r}(t)]^{\top}
+\bar{\nabla}^{\circ}\bar{r}(t)\cdot [\bar{\nabla}^{\circ}\bar{v}(t)]^{\top},
\]
where
\[ \bar{v}(t)=\frac{d\bar{r}}{dt}(t) \]
is the velocity vector for the network.
Bearing in mind that (Drozdov, 1996)
\[ \bar{\nabla}^{\circ}\bar{v}(t)=\bar{\nabla}^{\circ}\bar{r}(t)
\cdot \bar{\nabla}(t)\bar{v}(t), \]
where $\bar{\nabla}(t)$ is the gradient operator in the deformed state
at time $t$, we obtain
\begin{equation}
\frac{d\hat{C}}{dt}(t)=2\bar{\nabla}^{\circ}\bar{r}(t)\cdot\hat{D}(t)
\cdot [\bar{\nabla}^{\circ}\bar{r}(t)]^{\top},
\end{equation}
where
\[ \hat{D}(t)=\frac{1}{2}\Bigl [\bar{\nabla}(t)\bar{v}(t)
+(\bar{\nabla}(t)\bar{v}(t))^{\top}\Bigr ] \]
is the rate-of-strain tensor for the network.
Equation (20) implies that
\begin{equation}
\bar{l}\cdot \frac{d\hat{C}}{dt}(t)\cdot\bar{l}
= 2\bar{l}\cdot \bar{\nabla}^{\circ}\bar{r}(t)\cdot\hat{D}(t)
\cdot [\bar{\nabla}^{\circ}\bar{r}(t)]^{\top}\cdot \bar{l}
=2\hat{F}(t,\bar{l}):\hat{D}(t),
\end{equation}
where 
\begin{equation}
\hat{F}(t,\bar{l})=[\bar{\nabla}^{\circ}\bar{r}(t)]^{\top}\cdot
(\bar{l}\otimes \bar{l}) \cdot \bar{\nabla}^{\circ}\bar{r}(t)
\end{equation}
is the generalized Finger tensor,
the colon stands for convolution
and $\otimes$ denotes tensor product.
Substitution of Eqs. (19) and (21) into Eq. (17) results in the formula
\begin{equation}
\frac{dW}{dt}(t) =\hat{\Upsilon}(t):\hat{D}(t)+J_{0}(t)\frac{dT}{dt}(t)-J_{2}(t),
\end{equation}
where
\begin{eqnarray}
&& \hat{\Upsilon}(t) = \frac{\mu\Xi}{4\pi}\sum_{N=1}^{N_{1}}\frac{p_{N}}{N}
\int_{0}^{2\pi} d\varphi \int_{0}^{\pi}
\frac{e(t,\vartheta,\varphi)}{1-n(t,\vartheta,\varphi)}
\frac{\hat{F}(t,\vartheta,\varphi)}{\bar{l}\cdot\hat{C}(t)\cdot\bar{l}}
\sin\vartheta d\vartheta,
\nonumber\\
&& J_{2}(t) = \frac{\mu\Xi}{8\pi}\sum_{N=1}^{N_{1}}\frac{p_{N}}{N}
\int_{0}^{2\pi} d\varphi \int_{0}^{\pi}
\frac{h_{N}(t,\vartheta,\varphi)}{[1-n(t,\vartheta,\varphi)]^{2}}
\frac{\partial n}{\partial t}(t,\vartheta,\varphi)
\sin\vartheta d\vartheta,
\nonumber\\
&& h_{N}(t,\vartheta,\varphi) = e(t,\vartheta,\varphi)
\biggl \{ \frac{2\eta_{N} [1-n(t,\vartheta,\varphi)]}
{1+\eta_{N} n(t,\vartheta,\varphi)}
-e(t,\vartheta,\varphi) \biggr \}. 
\end{eqnarray}

\section{Constitutive equations}

Observations evidence that the growth of temperature driven by
periodic straining with small amplitudes (less than 10 \%)
is quite modest [it does not exceed 10 to 20 K for frequencies, $f$,
in the range from $10^{-1}$ to $10^{2}$ Hz, see Kar and Bhowmick (1997)
and Lion (1998)].
This implies that the absolute temperature, $T$, remains rather close 
to its reference value, $T_{0}$, and thermal expansion of the network 
may be disregarded.
Under the hypothesis about affine deformation (which states that 
the rate-of-strain tensor for deformation of a specimen coincides with 
the rate-of-strain tensor for the network, $\hat{D}$), the first 
law of thermodynamics for an incompressible network reads (Coleman and Gurtin, 1997)
\begin{equation}
\frac{d\Phi}{dt}=\frac{1}{\rho}\left (\hat{\sigma}_{\rm d}:\hat{D}
-\bar{\nabla}\cdot\bar{q} \right )+r, 
\end{equation}
where $\rho$ is mass density,
$\hat{\sigma}_{\rm d}$ is the deviatoric component
of the Cauchy stress tensor $\hat{\sigma}$,
$\bar{q}$ is the heat flux vector,
$\Phi$ is the internal energy and $r$ is the heat supply per unit mass.
At relatively small increments of temperature, the density of a polymer, 
$\rho$, is treated as a constant.
The Clausius--Duhem inequality is given by (Coleman and Gurtin, 1967)
\begin{equation}
\rho\frac{dQ}{dt}=\rho\frac{dS}{dt}
+\bar{\nabla} \cdot \Bigl (\frac{\bar{q}}{T}\Bigr )
-\frac{\rho r}{T} \geq 0, 
\end{equation}
where $S$ is the entropy and $Q$ is the entropy production per unit mass.
Bearing in mind that
\[ \bar{\nabla}\cdot\Bigl (\frac{\bar{q}}{T}\Bigr )
=\frac{1}{T}\bar{\nabla}\cdot\bar{q}-\frac{1}{T^{2}}
\bar{q}\cdot\bar{\nabla} T \]
and excluding the term $\bar{\nabla}\cdot\bar{q}$ from Eqs. (25) and (26), we find that
\begin{equation}
T \frac{dQ}{dt}=T\frac{dS}{dt}-\frac{d\Phi}{dt}
+\frac{1}{\rho}\Bigl (\hat{\sigma}_{\rm d}:\hat{D}
-\frac{1}{T}\bar{q}\cdot \bar{\nabla} T \Bigr ) \geq 0.
\end{equation}
The internal energy per unit mass, $\Phi$, is connected 
with the free (Helmholtz) energy per unit mass, $\Psi$, by the equality
\begin{equation}
\Phi=\Psi+ST.
\end{equation}
It follows from Eqs. (27) and (28) that
\begin{equation}
T \frac{dQ}{dt}=-S\frac{dT}{dt}-\frac{d\Psi}{dt}
+\frac{1}{\rho}\Bigl ( \hat{\sigma}_{\rm d}:\hat{D}
-\frac{1}{T}\bar{q}\cdot \bar{\nabla} T \Bigr) \geq 0.
\end{equation}
It is assumed that the free energy, $\Psi$, consists of the following four terms:
\begin{enumerate}
\item
the free energy, $\Psi_{\rm eq}$, of the network
in the equilibrium stress-free state at the reference temperature $T_{0}$
(when the numbers of RHMs in a strand coincides with its initial value, $N$);

\item
the free energy of thermal motion, $\Psi_{\rm th}$,
which is determined by the conventional formula
\[ \Psi_{\rm th}=(c-S_{0})(T-T_{0}) -cT\ln\frac{T}{T_{0}}, \]
where $c$ is the specific heat (at relatively small increments of temperature, $c$ is
treated as a constant) and $S_{0}$ is the entropy per unit mass in the stress-free 
state at the reference temperature $T_{0}$;

\item
the potential energy of deformation for the network, $W$, given by Eq. (16);

\item
the increment of free energy, $\Delta \Psi$, induced by suppression and thermal
activation of RHMs.
\end{enumerate}
The latter term is associated with the excess free energy of a strand
with the current number of RHMs, $N_{\rm a}$, over that for the strand
with the equilibrium number of RHMs, $N$.
As a first approximation, the increment of the free energy for a strand with $N$ RHMs, 
$\Delta \psi_{N}$, is given by
\begin{equation}
\Delta \psi_{N}=\frac{1}{2}\beta \mu ( N_{\rm a}-N) ^{2}, 
\end{equation}
where $\beta$ is a dimensionless parameter (which, in general, depends on temperature),
and the average rigidity per RHM, $\mu$, is introduced to ensure
the correct dimension of the function $\Delta \psi_{N}$.
Summing the increments of free energy, Eq. (30), for strands with various numbers 
of RHMs, $N$, and various guiding vectors, $\bar{l}$, 
and using Eqs. (3) and (15), we find that
\[ \Delta \Psi (t)=\frac{\beta\mu \Xi}{8\pi}\sum_{N=1}^{N_{1}} N^{2} p_{N}
\int_{0}^{2\pi}d\varphi\int_{0}^{\pi}
n^{2}(t,\vartheta,\varphi)\sin\vartheta d\vartheta. \]
The free energy per unit mass of the network, $\Psi$, reads
\begin{equation}
\Psi = \Psi_{\rm eq}+ (c-S_{0})(T-T_{0}) -cT\ln\frac{T}{T_{0}}+W
+ \frac{\beta\mu \Xi}{8\pi} \sum_{N=1}^{N_{1}} N^{2} p_{N}
\int_{0}^{2\pi}d\varphi\int_{0}^{\pi} n^{2}\sin\vartheta d\vartheta.
\end{equation}
It follows from Eqs. (23), (24) and (31) that the derivative of the free energy per unit mass,
$\Psi$, with respect to time, $t$, is given by
\begin{equation}
\frac{d\Psi}{dt} = \hat{\Upsilon}:\hat{D}+\Bigl (\Delta S-S_{0}-c\ln\frac{T}{T_{0}}\Bigr )\frac{dT}{dt}-J,
\end{equation}
where
\begin{eqnarray}
\Delta S(t) &=& J_{0}(t)+\frac{\Xi}{8\pi} \frac{d(\beta\mu)}{dT} \sum_{N=1}^{N_{1}} N^{2} p_{N}
\int_{0}^{2\pi}d\varphi\int_{0}^{\pi} n^{2}\sin\vartheta d\vartheta,
\nonumber\\
J(t) &=& \frac{\mu\Xi}{8\pi} \sum_{N=1}^{N_{1}} \frac{p_{N}}{N}
\int_{0}^{2\pi} d\varphi \int_{0}^{\pi}
\frac{H_{N} (t,\vartheta,\varphi)}{[1-n(t,\vartheta,\varphi)]^{2}}
\frac{\partial n}{\partial t}(t,\vartheta,\varphi)
\sin\vartheta d\vartheta,
\nonumber\\
H_{N} (t,\vartheta,\varphi) &=& e(t,\vartheta,\varphi)
\biggl \{ \frac{2\eta_{N} [1-n(t,\vartheta,\varphi)]}{1+\eta_{N}n(t,\vartheta,\varphi)}
-e(t,\vartheta,\varphi) \biggr \}
-2\beta N^{3} n(t,\vartheta,\varphi) [1-n(t,\vartheta,\varphi)]^{2}.
\end{eqnarray}
Combining Eqs. (29) and (32), we find that
\begin{equation}
T\frac{dQ}{dt}=\frac{1}{\rho}\Bigl (\hat{\sigma}_{\rm d}-\rho\hat{\Upsilon}\Bigr ):\hat{D}
-\Bigl (S-S_{0}-c\ln\frac{T}{T_{0}}+\Delta S\Bigr )\frac{dT}{dt}
-\frac{1}{\rho T}\bar{q}\cdot\bar{\nabla}T+ J\geq 0.
\end{equation}
Applying the conventional reasoning to Eq. (34), see Coleman and Gurtin (1967),
we find that the expressions in braces vanish.
This implies the formula for the entropy per unit mass, 
\begin{equation}
S=S_{0}+c\ln\frac{T}{T_{0}}-\Delta S
\end{equation}
and the relationship for the Cauchy stress tensor
\begin{equation}
\hat{\sigma}(t) = -P(t)\hat{I} + \frac{\rho \mu\Xi}{4\pi}
\sum_{N=1}^{N_{1}} \frac{p_{N}}{N} \int_{0}^{2\pi} d\varphi \int_{0}^{\pi}
\frac{e(t,\vartheta,\varphi)}{1 -n(t,\vartheta,\varphi)}
\frac{\hat{F}(t,\vartheta,\varphi)}{\bar{l}\cdot\hat{C}(t)\cdot\bar{l}}
\sin\vartheta d\vartheta,
\end{equation}
where $P(t)$ is pressure.
It follows from Eq. (35) that the quantity $\Delta S$ [given by Eqs. (18) and (33)]
characterizes changes in the entropy per unit mass, $S$, driven by deformation
of the network and suppression--activation of RHMs.
Substitution of Eq. (22) into Eq. (36) results in the constitutive equation
\begin{equation}
\hat{\sigma}(t) = -P(t)\hat{I}+[\bar{\nabla}^{\circ}\bar{r}(t)]^{\top}\cdot
\hat{B}(t) \cdot \bar{\nabla}^{\circ}\bar{r}(t)
\end{equation}
with
\begin{equation}
\hat{B}(t)= \frac{\rho \mu\Xi}{4\pi} \sum_{N=1}^{N_{1}} \frac{p_{N}}{N}
\int_{0}^{2\pi} d\varphi \int_{0}^{\pi}
\frac{e(t,\vartheta,\varphi)}{1-n(t,\vartheta,\varphi)}
\frac{\bar{l}\otimes \bar{l}}{\bar{l} \cdot\hat{C}(t)\cdot \bar{l}}
\sin\vartheta d\vartheta.
\end{equation}
Combining Eqs. (34) to (36), we obtain
\begin{equation}
T\frac{dQ}{dt}=J-\frac{1}{\rho T}\bar{q}\cdot\bar{\nabla}T \geq 0.
\end{equation}
The function $J(t)$ determines the rate of entropy production (per unit mass)
induced by changes in the numbers of RHMs, whereas the last term on
the right-hand side of Eq. (39) characterizes entropy production (per unit mass
and unit time) driven by thermal conductivity.
The function $Q$ is assumed to be presented in the form
\begin{equation}
Q(t)=Q_{\rm th}(t)+\Delta Q(t),
\end{equation}
where $Q_{\rm th}$ describes the entropy production induced by thermal
diffusivity and $\Delta Q$ determines the entropy production associated with 
the suppression--activation mechanism.
We accept the conventional formula for the derivative of the function $Q_{\rm th}$,
\begin{equation}
\frac{dQ_{\rm th}}{dt}(t)=\frac{\kappa (\bar{\nabla}T)^{2}}{\rho T^{2}},
\end{equation}
where $\kappa>0$ is a constant thermal diffusivity, and 
$(\bar{\nabla}T)^{2}=\bar{\nabla}T\cdot\bar{\nabla}T$.
It follows from Eqs. (39) to (41) that
\begin{equation}
T\frac{d\Delta Q}{dt}=J-\frac{1}{\rho T}
\Bigl (\bar{q}+\kappa\bar{\nabla} T\Bigr )\cdot \bar{\nabla} T\geq 0.
\end{equation}
Bearing in mind that this inequality holds for an arbitrary thermo-mechanical
process, we find that the expression in braces vanishes, which results in the Fourier law
for the heat flux vector 
\begin{equation}
\bar{q}(t)=-\kappa \bar{\nabla}(t)T(t). 
\end{equation}
Equations (42) and (43) yield
\begin{equation}
T\frac{d\Delta Q}{dt}=J\geq 0.
\end{equation}
We postulate that the function $\Delta Q$ is given by
\begin{equation}
\Delta Q(t)=\frac{1}{4\pi} \sum_{N=1}^{N_{1}}\Xi_{N}\int_{0}^{2\pi} d\varphi 
\int_{0}^{\pi} \Delta Q_{N}(t,\vartheta,\varphi)\sin\vartheta d\vartheta,
\end{equation}
where $\Delta Q_{N}(t,\bar{l})$ characterizes the energy dissipation per strand 
containing $N$ RHMs and directed along vector $\bar{l}$.
To construct the functions $\Delta Q_{N}$, we refer to the conventional expression
for the entropy production $\tilde{Q}$ in an incompressible viscous fluid
\begin{equation}
T\frac{d\tilde{Q}}{dt}=\zeta \hat{D}:\hat{D},
\end{equation}
where $\zeta$ stands for viscosity.
Equation (46) means that the rate of entropy production per unit mass is proportional
to the square of the relative velocity (which is characterized by the rate-of-strain
tensor $\hat{D}$).
By analogy with this relation, we postulate that the expression
\[ T(t) \frac{\partial \Delta Q_{N}}{\partial t}(t,\vartheta,\varphi) \]
is proportional to the square of the relative rate of changes in the number of RHMs.
According to Eq. (3), this rate is given by
\begin{equation}
-\frac{1}{N_{\rm a}}\frac{dN_{\rm a}}{dt}= \frac{1}{1-n}\frac{dn}{dt}. 
\end{equation}
With reference to Clarke et al. (2000), we assume that energy can dissipate
at junction points only, which implies that Eq. (47) should be multiplied by
the probability that the activated (suppressed) segment is located at the
end of a strand (this probability equals $2/N$ for a strand containing $N$ RHMs).
As a result, we obtain
\begin{equation}
T(t)\frac{\partial \Delta Q_{N}}{\partial t}(t,\vartheta,\varphi)=\frac{\tilde{\kappa}}{N^{2}}
\Bigl [ 1-n(t,\vartheta,\varphi) \Bigr ]^{-2}
\Bigl [\frac{\partial n}{\partial t} (t,\vartheta,\varphi)\Bigr ]^{2},
\end{equation}
where $\tilde{\kappa}$ is a positive coefficient independent of the number of
RHMs in a strand.
Substituting expressions (45) and (48) into Eq. (44) and equating coefficients at
the derivative $\partial n/\partial t$ on the left and right hand sides on
the resulting equality, we arrive at the kinetic equation 
for the function $n(t,\vartheta,\varphi)$, 
\[ \frac{\partial n}{\partial t}(t,\vartheta,\varphi)
= \frac{\mu N}{2\tilde{\kappa}} H_{N}(t,\vartheta,\varphi). \]
It follows from this formula and Eq. (33) that
\begin{equation}
\frac{\partial n}{\partial t}(t,\vartheta,\varphi)
= \alpha_{1N} e(t,\vartheta,\varphi)
\biggl \{ \frac{2\eta_{N} [1-n(t,\vartheta,\varphi)]}
{1+\eta_{N} n(t,\vartheta,\varphi)} -e(t,\vartheta,\varphi)\biggr \}
-\alpha_{2N}n(t,\vartheta,\varphi) [1-n(t,\vartheta,\varphi)]^{2},
\end{equation}
where
\begin{equation}
\alpha_{1N}=a_{1}N,
\qquad
\alpha_{2N}=a_{2}N^{4},
\qquad
a_{1}=\frac{\mu}{2\tilde{\kappa}},
\qquad
a_{2}=\frac{\mu \beta}{\tilde{\kappa}}.
\end{equation}
It follows from Eq. (3) that the initial condition for Eq. (49) reads
\begin{equation}
n(t,\vartheta,\varphi)\Bigl |_{t=0}=0.
\end{equation}
Relations (13), (37), (38), (49) and (51) provide a set of constitutive equations for
a rubbery polymer at finite strains.
The behavior of the coefficients (50) in kinetic equation (49) is unusual.
The parameter $\alpha_{1N}$ linearly increases with $N$, which means that suppression
of RHMs occurs more rapidly in long strands than in short ones.
On the other hand, the quantity $\alpha_{2N}$ grows as the fourth power of $N$,
which implies that the concentration of RHMs in long strands remains sufficiently large.

\section{Constitutive equations at small strains}

We now simplify the constitutive equations for the case of small strains.
According to Eq. (49), the condition $|e|\ll 1$ ensures small changes 
in the concentrations of RHMs, $n\ll 1$.
Introducing the strain tensor $\hat{\epsilon}$ by the conventional formula
\[ \hat{\epsilon}(t)=\frac{1}{2}[\hat{C}(t)-\hat{I}], \]
we find from Eq. (13) that
\[ e(t,\bar{l})=\frac{1}{2}\ln \Bigl [
1+2\bar{l}\cdot\hat{\epsilon}(t)\cdot\bar{l}\Bigr ]
-\ln \Bigl [ 1+\eta_{N} n(t,\bar{l}) \Bigr ]. \]
Neglecting terms beyond the first order of smallness compared 
to $|\hat{\epsilon}|$ and $n$, we obtain
\begin{equation}
e(t,\bar{l})=\bar{l}\cdot\hat{\epsilon}(t)\cdot\bar{l}
-\eta_{N} n(t,\bar{l}).
\end{equation}
It follows from Eqs. (49) and (52) that with the required level of accuracy,
$n$ satisfies the differential equation
\begin{equation}
\frac{\partial n}{\partial t}(t,\bar{l})
=2\alpha_{1N}\eta_{N} \bar{l}\cdot\hat{\epsilon}(t)\cdot\bar{l}
-(2\alpha_{1N}\eta_{N}^{2}+\alpha_{2N}) n(t,\bar{l}).
\end{equation}
The solution of Eq. (53) with initial condition (51) reads
\[ n(t,\bar{l})=2\alpha_{1N}\eta_{N} \int_{0}^{t} \exp \Bigl [ -
(2\alpha_{1N}\eta_{N}^{2}+\alpha_{2N}) (t-\tau) \Bigr ]
\bar{l}\cdot\hat{\epsilon} (\tau)\cdot\bar{l}\;\; d\tau. \]
This equality together with Eq. (52) implies that
\begin{equation}
e(t,\bar{l}) = \bar{l}\cdot \biggl \{ \hat{\epsilon}(t)
-2\alpha_{1N}\eta_{N}^{2} \int_{0}^{t} \exp \Bigl [ -
(2\alpha_{1N}\eta_{N}^{2}+\alpha_{2N}) (t-\tau) \Bigr ]
\hat{\epsilon} (\tau)\; d\tau\biggr \} \cdot\bar{l}.
\end{equation}
Substituting Eq. (52) into Eqs. (37) and (38) and neglecting terms beyond
the first order of smallness, we find that
\begin{eqnarray}
\hat{\sigma}(t) &=& -P(t)\hat{I} +\frac{\rho \mu\Xi}{4\pi}\sum_{N=1}^{N_{1}}
\frac{p_{N}}{N} \int_{0}^{2\pi} d\varphi \int_{0}^{\pi} (\bar{l}\otimes\bar{l})
\sin\vartheta d\vartheta
\nonumber\\
&& \bar{l}\cdot \biggl \{ \hat{\epsilon}(t)
-2\alpha_{1N}\eta_{N}^{2} \int_{0}^{t} \exp \Bigl [ -
(2\alpha_{1N}\eta_{N}^{2}+\alpha_{2N}) (t-\tau) \Bigr ]
\hat{\epsilon} (\tau)\; d\tau\biggr \} \cdot\bar{l}.
\end{eqnarray}
To calculate the integral, we introduce a Cartesian coordinate
frame $\{ x_{i} \}$ with base vectors $\bar{e}_{i}$ which are directed
along the eigenvectors of the symmetric tensor $\hat{\epsilon}$
and obtain
\begin{equation}
\hat{\epsilon}(t)=\epsilon_{1}(t)\bar{e}_{1}\bar{e}_{1}
+\epsilon_{2}(t)\bar{e}_{2}\bar{e}_{2}
+\epsilon_{3}(t)\bar{e}_{3}\bar{e}_{3}, 
\end{equation}
where $\epsilon_{i}$ is the $i$th eigenvalue of $\hat{\epsilon}$.
The unit vector $\bar{l}$ is given by
\begin{equation}
\bar{l}=\cos\vartheta \bar{e}_{1}+
\sin\vartheta (\cos\varphi \bar{e}_{2}+\sin\varphi \bar{e}_{3} ),
\end{equation}
It follows from Eqs. (56) and (57) that
\[ \bar{l}\cdot\hat{\epsilon}(t)\cdot\bar{l}=
\epsilon_{1}(t)\cos^{2}\vartheta+\Bigl [\epsilon_{2}(t)\cos^{2}\varphi
+\epsilon_{3}(t)\sin^{2}\varphi \Bigr ]\sin^{2}\vartheta. \]
Equation (57) implies that the tensor $\bar{l}\otimes \bar{l}$
is determined by the matrix
\[ \bar{l}\otimes\bar{l}=\left [\begin{array}{ccc}
\cos^{2}\vartheta & \sin\vartheta\cos\vartheta\cos\varphi &
\sin\vartheta\cos\vartheta\sin\varphi\\
\sin\vartheta\cos\vartheta\cos\varphi & \sin^{2}\vartheta\cos^{2}\varphi &
\sin^{2}\vartheta\sin\varphi\cos\varphi\\
\sin\vartheta\cos\vartheta\sin\varphi & \sin^{2}\vartheta\sin\varphi\cos\varphi
& \sin^{2}\vartheta\sin^{2}\varphi
\end{array}\right ]. \]
After simple algebra we find that
\[ \frac{1}{4\pi}\int_{0}^{2\pi}d\varphi
\int_{0}^{\pi} ( \bar{l}\otimes \bar{l}) \sin\vartheta d\vartheta\;
\bar{l}\cdot\hat{\epsilon}\cdot\bar{l}
=\frac{1}{15}\left [\begin{array}{ccc}
3\epsilon_{1}+\epsilon_{2}+\epsilon_{3} & 0 & 0\\
0 & \epsilon_{1}+3\epsilon_{2}+\epsilon_{3} & 0\\
0 & 0 & \epsilon_{1}+\epsilon_{2}+3\epsilon_{3}
\end{array}\right ]. \]
This equality together with the incompressibility condition,
\[ \epsilon_{1}+\epsilon_{2}+\epsilon_{3}=0, \]
results in the formula
\[ \frac{1}{4\pi}\int_{0}^{2\pi}d\varphi\int_{0}^{\pi}
(\bar{l}\otimes \bar{l}) \sin\vartheta d\vartheta\;
\bar{l}\cdot\hat{\epsilon}\cdot\bar{l}
=\frac{2}{15}\hat{\epsilon}. \]
Substituting this expression into Eq. (55), we arrive at the stress--strain relation
\begin{equation}
\hat{\sigma}(t) = -P(t)\hat{I} +\frac{2\rho \mu\Xi}{15}\sum_{N=1}^{N_{1}}
\frac{p_{N}}{N} \biggl [ \hat{\epsilon}(t)
-\zeta_{N}\Gamma_{N} \int_{0}^{t} \exp \Bigl ( -\Gamma_{N}(t-\tau) \Bigr )
\hat{\epsilon} (\tau)d\tau\biggr ]
\end{equation}
with
\begin{equation}
\Gamma_{N}=2\alpha_{1N}\eta_{N}^{2}+\alpha_{2N},
\qquad
\zeta_{N}=\frac{2\alpha_{1N}\eta_{N}^{2}}{2\alpha_{1N}\eta_{N}^{2}+\alpha_{2N}}. 
\end{equation}
It follows from Eq. (58) that the conventional shear modulus of a rubbery polymer
at small strains is given by
\[ G_{0}=\frac{\rho \mu \Xi}{15}\sum_{N=1}^{N_{1}} \frac{p_{N}}{N}.  \]
This equation establishes correspondence between the material parameters at the micro-level
($\mu$ and $N$) and those at the macro-level ($G_{0}$ and $\rho$).
Formula (58) is equivalent to phenomenological relations (1) and (2).
An important advantage of Eq. (58) compared to those equations is that 
the relaxation rates $\Gamma_{N}$ [inverse to the characteristic times $\tau_{N}$ is Eq. (2)] 
and the coefficients $\zeta_{N}$ [analogs of the relaxation strengths $\xi_{N}$ is Eq. (2)] 
are not arbitrary, but are determined by Eq. (59).

\section{Uniaxial extension of a specimen}

We consider a specimen in the form of a rectilinear bar whose points 
refer to Cartesian coordinates $\{x_{i}\}$ with unit vectors $\bar{e}_{i}$.
At uniaxial extension of an incompressible material, the strain tensor $\hat{\epsilon}$ 
is given by
\begin{equation}
\hat{\epsilon}=\epsilon \Bigl [\bar{e}_{1}\bar{e}_{1}-\frac{1}{2}
(\bar{e}_{2}\bar{e}_{2}+\bar{e}_{3}\bar{e}_{3})\Bigr ],
\end{equation}
where $\epsilon$ is the longitudinal strain.
Substitution of Eq. (60) into Eq. (58) implies that the stress tensor reads
\[ \hat{\sigma}=\sigma \bar{e}_{1}\bar{e}_{1}+\sigma_{0}
(\bar{e}_{2}\bar{e}_{2}+\bar{e}_{3}\bar{e}_{3}), \]
where
\begin{eqnarray}
\sigma(t) &=& -P(t)+\frac{2\rho \mu\Xi}{15}\sum_{N=1}^{N_{1}}
\frac{p_{N}}{N} \biggl [ \epsilon(t)
-\zeta_{N}\Gamma_{N} \int_{0}^{t} \exp \Bigl ( -\Gamma_{N}(t-\tau) \Bigr )
\epsilon (\tau)d\tau\biggr ],
\nonumber\\
\sigma_{0}(t) &=& -P(t)-\frac{\rho \mu\Xi}{15}\sum_{N=1}^{N_{1}}
\frac{p_{N}}{N} \biggl [ \epsilon(t)
-\zeta_{N}\Gamma_{N} \int_{0}^{t} \exp \Bigl ( -\Gamma_{N}(t-\tau) \Bigr )
\epsilon (\tau)d\tau\biggr ].
\end{eqnarray}
It follows from the boundary condition on the lateral surface of
the specimen that $\sigma_{0}(t)=0$.
Combining this equality with Eq. (61), we arrive at the formula for
the longitudinal stress
\begin{equation}
\sigma(t)=K \sum_{N=1}^{N_{1}}
\frac{p_{N}}{N} \biggl [ \epsilon(t)
-\zeta_{N}\Gamma_{N} \int_{0}^{t} \exp \Bigl ( -\Gamma_{N}(t-\tau) \Bigr )
\epsilon (\tau)d\tau\biggr ]
\end{equation}
with
\[ K=\frac{\rho \mu\Xi}{5}. \]
The parameter $K$ is proportional to Young's modulus, $E_{0}$, which is determined by the formula
\begin{equation}
E_{0}=K \sum_{N=1}^{N_{1}} \frac{p_{N}}{N}. 
\end{equation}
To study the response of a specimen to periodic stretching, we set
\begin{equation}
\epsilon(t)=\Delta \exp (\imath\omega t),
\end{equation}
where $\imath=\sqrt{-1}$, $\Delta$ is the amplitude and $\omega$ is the frequency of
oscillations,
substitute expression (64) into Eq. (62), and obtain
\begin{eqnarray}
\sigma(t) &=& K \sum_{N=1}^{N_{1}}
\frac{p_{N}}{N} \biggl \{ 1 -\zeta_{N}\Gamma_{N} 
\int_{0}^{t} \exp \Bigl [ -(\Gamma_{N}+\imath\omega)(t-\tau) \Bigr ]
d\tau\biggr \}\epsilon(t)
\nonumber\\
&=& K\sum_{N=1}^{N_{1}} \frac{p_{N}}{N} \biggl \{ 1 -\zeta_{N}\Gamma_{N} 
\int_{0}^{t} \exp \Bigl [ -(\Gamma_{N}+\imath\omega)s \Bigr ] ds \biggr \}\epsilon(t).
\end{eqnarray}
Adopting the conventional approach to the analysis of steady oscillations,
we replace the upper limit of integration in Eq. (65) by infinity, 
calculate the integral, introduce the complex modulus
\[ E^{\ast}=\frac{\sigma}{\epsilon}, \]
and find that
\[ E^{\ast}(\omega)=K \sum_{N=1}^{N_{1}} \frac{p_{N}}{N} 
\biggl (1-\frac{\zeta_{N}\Gamma_{N}}{\Gamma_{N}+\imath\omega}\biggr ). \]
Bearing in mind that
\[ E^{\ast}=E^{\prime}+\imath E^{\prime\prime}, \]
where $E^{\prime}$ is the storage modulus and
$E^{\prime\prime}$ is the loss modulus, we arrive at the formulas
\begin{equation}
E^{\prime}(\omega)=K\sum_{N=1}^{N_{1}}
\frac{p_{N}}{N} \frac{(1-\zeta_{N})\Gamma_{N}^{2}+\omega^{2}}{\Gamma_{N}^{2}+\omega^{2}},
\qquad
E^{\prime\prime}(\omega)=K \sum_{N=1}^{N_{1}}
\frac{p_{N}}{N} \frac{\zeta_{N}\Gamma_{N}\omega}{\Gamma_{N}^{2}+\omega^{2}}.
\end{equation}
Equations (66) are rather standard (except for the presence of the multiplier
$p_{N}/N$) and may be found in textbooks on the mechanics of polymers (see, e.g., Ferry, 1980).
The novelty of our approach is that it provides a new scaling for the relaxation rates, 
$\Gamma_{N}$, and the relaxation strengths, $\zeta_{N}$.
It follows from Eqs. (12), (50) and (59) that
\begin{equation}
\Gamma_{N}=aN^{2}+bN^{4},
\qquad
\zeta_{N}=\frac{a}{a+bN^{2}},
\end{equation}
where
\begin{equation}
a=2a_{1}\tilde{\eta}^{2}=\frac{\mu}{\tilde{\kappa}}\tilde{\eta}^{2},
\qquad
b=a_{2}=\frac{\mu}{\tilde{\kappa}}\beta.
\end{equation}
The advantage of Eqs. (14), (66) and (67) is that they contain 
only 5 adjustable parameters (which, in general, are temperature-dependent): 
the elastic modulus $K$,
the average number of RHMs per strand in the equilibrium state $\tilde{N}$,
the standard deviation of the number of RHMs $\Sigma$,
the rate of suppression of RHMs $a$,
and the rate of activation for frozen segments $b$
(the parameter $N_{1}$ is chosen in such a way that $p_{N_{1}}$ becomes negligible
compared to unity; for the quasi-Gaussian distribution (14),  it suffices to 
set $N_{1}>\tilde{N}+3\Sigma$).
This number is essentially less than the number of adjustable constants
conventionally employed to fit experimental data [for example, we refer to Wu and
Liechti (2000), where 14 constants in Eq. (2), $\tau_{N}$, were fixed {\it a priori}
and the other 14 constants, $\xi_{N}$, were found by the least-square algorithm].

\section{Comparison with experimental data}

To verify the constitutive equations (14), (66) and (67),
we fit experimental data for a carbon black filled natural rubber vulcanizate
with the filler content 50 phr (parts per hundred parts of rubber)
and the glass transition temperature $T_{\rm g}=220$ K.
For a detailed description of specimens and the experimental procedure,
the reader is referred to Lion (1998).

We begin with matching experimental data in dynamic tensile tests
at the temperature $T=253$ K, the smallest amplitude of oscillations
$\Delta=0.006$ and the frequency $f=\omega/(2\pi)$ in the range from $10^{-1}$
to $10^{2}$ Hz.
Given a modulus $K$, the quantities $a$, $b$, $\tilde{N}$ and $\Sigma$ that
ensure the best quality of approximation for the storage modulus, $E^{\prime}$,
are found by the steepest-descent procedure.
The parameter $K$ is determined by the least-squares algorithm.
Afterwards, we fix the constants $a$ and $b$ found by matching observations
for $E^{\prime}(\omega)$ with $\Delta=0.006$ and fit experimental data for other
amplitudes of oscillations, $\Delta=0.011$, $\Delta=0.028$ and $\Delta=0.056$,
by using three adjustable parameters: $K$, $\tilde{N}$ and $\Sigma$.
Experimental data and results of numerical analysis (with $N_{1}=500$)
are plotted in Figure 2.
The same procedure of fitting is repeated for three other temperatures: $T=296$,
$T=333$ and $T=373$ K.
Appropriate results are depicted in Figures 3 to 5.
Figures 2 to 5 demonstrate fair agreement between observations and results 
of numerical simulation at all temperatures under consideration.

The average number of RHMs in an equilibrated strand, $\tilde{N}$, is plotted versus
the amplitude of oscillations, $\Delta$, in Figures 7 to 10.
Experimental data are approximated by the phenomenological relation
\begin{equation}
\tilde{N}=\tilde{N}_{0}+\tilde{N}_{1}\Delta,
\end{equation}
where the constants $\tilde{N}_{i}$ are found by the least-squares technique.
Figures 6 to 9 show that Eq. (69) correctly predicts observations in the entire
rang of amplitudes, except for a region of very small strains.
The latter may be explained by insufficient accuracy of the dynamic tests
with extremely small amplitudes on preloaded specimens (with the compression strain of 
about 10 \%).

The standard deviation of the number of RHMs in a strand, $\Sigma$,
is depicted in Figures 10 to 13 as a function of the amplitude of oscillations,
$\Delta$.
Observations are approximated by the linear dependence
\begin{equation}
\Sigma=\Sigma_{0}+\Sigma_{1}\Delta,
\end{equation}
where adjustable parameters $\Sigma_{i}$ are determined by using the least-squares
algorithm.
Equation (70) provides an acceptable quality of matching observations in the
entire region of amplitudes of oscillations, $\Delta$, except for very low intencities
of strains.

The parameters $\tilde{N}_{i}$ and $\Sigma_{i}$ found at various temperatures are 
collected in Table 1 which demonstrates that the average
number of RHMs per strand grows with the amplitude of straining (in accord with
the intuitive picture that mechanical loading increases the molecular mobility of chains).

The quantities $a$ and $b$ are plotted versus temperature $T$ in Figure 14.
Because the suppression--activation mechanism for RHMs is assumed to be driven 
by thermal fluctuations, 
the rates of transformation are approximated by the Arrhenius dependences,
\[ a=\tilde{a}\exp\Bigl (-\frac{E_{a}}{RT}\Bigr ),
\qquad
b=\tilde{b}\exp\Bigl (-\frac{E_{b}}{RT}\Bigr ),
\]
where $\tilde{a}$ and $\tilde{b}$ are appropriate rates at high temperature,
$E_{a}$ and $E_{b}$ are activation energies,
and $R$ is the universal gas constant.
It follows from these formulas that
\begin{equation}
\log a=a_{0}-\frac{a_{1}}{T},
\qquad
\log b=b_{0}-\frac{b_{1}}{T},
\end{equation}
where
\[ a_{0}=\log \tilde{a},
\qquad
a_{1}=\frac{E_{a}}{R\ln 10},
\qquad
b_{0}=\log \tilde{b},
\qquad
b_{1}=\frac{E_{b}}{R\ln 10},
\qquad
\log=\log_{10}. \]
Figure 14 demonstrates an acceptable agreement between experimental data and their
approximation by Eqs. (71) with adjustable parameters $a_{i}$ and $b_{i}$, except for
the region of high temperatures, where relations (71) overestimate the transformation rates.
The latter may be explained by the following reasons:
\begin{enumerate}
\item
the average radius of a tube which reflects constrains on the motion of a strand 
imposed by surrounding macromolecules strongly increases at elevated temperatures,
which implies that transformations of RHMs are slowed down;

\item
equilibration of an elastomer specimen at temperatures near $T=100$ $^{\circ}$C 
before mechanical tests results in a noticeable scission of chains 
(see, e.g., Palmas et al., 2001), which is beyond the scope of the present model.
\end{enumerate}

The parameter $b$ grows with temperature faster than $a$
($b_{1}=2.32\cdot 10^{3}$ versus $a_{1}=1.48\cdot 10^{3}$).
Assuming the quantity $\tilde{\eta}$ to depend on temperature rather weakly
(as a power law, but not in the exponential fashion),
we find from Eq. (68) that $\beta$ substantially increases with temperature.
The latter means that the excess free energy caused by suppression and
activation of RHMs, see Eq. (30), is a strongly increasing function of temperature.

Another important conclusion may be drawn from results listed in Table 1
which show that the parameters $\tilde{N}_{1}$ and $\Sigma_{1}$ increase
with temperature (the only exception is the value of ${N}_{1}$ at the lowest
temperature, $T=253$ K, when the elastomer demonstrates quasi-brittle response).
This implies that the influence of the amplitude of oscillations, $\Delta$, 
on the distribution of strands with various numbers of RHMs becomes stronger 
with temperature.

It is worth noting that the parameters $\tilde{N}$ and $\Sigma$
corresponding to the smallest amplitude of straining, $\Delta=0.006$,
are practically independent of temperature, see data collected in Table 2.
This means that mechanical loading may be thought of as the main reason for changes 
in the distribution of RHMs.
The effect of mechanical factors is enchanced at elevated temperatures
and is weakened at relatively low temperatures, in agreement with the assumption
about the entropic nature of transformations of RHMs.

\section{Discussion}

Our purpose now is to develop conditions under which the viscoelastic response of
rubbery polymers is substantially weakened 
and to establish some correspondence between these relations
and observations exposed in the introductory section.

It follows from Eqs. (67) and (68) that
\begin{equation}
\zeta_{N}=\Bigl (1+\frac{\beta}{\tilde{\eta}^{2}}N^{2}\Bigr )^{-1}.
\end{equation}
Combining Eqs. (12), (66) and (72), replacing the parameter $N$ by its average value, $\tilde{N}$
and assuming $\Gamma_{N}$ to be of the same order of magnitude as the frequency of oscillations
(which results in a very rough estimate), we conclude that the mechanical damping
of an elastomer disappears ($E^{\prime\prime}\to 0$), provided that
\begin{equation}
\beta\tilde{N}^{2}\Bigl ( \frac{\tilde{\delta}}{\delta^{\circ}}\Bigr )^{2}\gg 1.
\end{equation}
Based on Eq. (73) one can speculate that the weak damping reported by Leblanc and
Cartault (2001) is associated with a large average number of RHMs per strand, $\tilde{N}$, 
in uncured styrene--butadiene composites.
This conclusion does not contradict the presence of filler particles in these compounds, 
because at the length scale of RHMs (which is less than the length-scale of a strand 
by two orders of magnitude, see Table 1), the influence of particles is negligible.

Observations by Liang et al. (1999) may be explained within the framework of Eq. (73)
by the large energy necessary for suppression of a RHM, $\beta$, in a polymer composite
with a high degree of crystallinity and a relatively small content of elastomer (about 10 \%).

To explain experimental results reported by Aksel and H\"{u}bner (1996)
and Clarke et al. (2000), we consider uniaxial relaxation test with a small strain $\epsilon_{0}$,
\begin{equation}
\epsilon(t)=0\quad
(t<0),\qquad
\epsilon(t)=\epsilon_{0}
\quad
(t>0),
\end{equation}
substitute expression (74) into Eq. (62), introduce the relaxation modulus 
\[E(t)=\frac{\sigma(t)}{\epsilon_{0}}, \]
and find that
\begin{equation}
E(t)=K\sum_{N=1}^{N_{1}} \frac{p_{N}}{N}
\Bigl [ 1-\zeta_{N}\Bigl (1-\exp (-\Gamma_{N}t)\Bigr )\Bigr ].
\end{equation}
It follows from Eq. (75) that changes in the relaxation modulus, $E(t)$, are negligible
when either the parameters $\zeta_{N}$ are small compared to unity
[which is described by Eq. (73)] or the characteristic times of relaxation
$\tau_{N}=\Gamma_{N}^{-1}$ are small compared to the characteristic time of the test, $\tau_{0}$.
Equations (12), (67) and (68) imply that the latter condition 
is given by
\[ \frac{\tilde{\kappa}}{\mu N^{2}}\Bigl (\frac{\tilde{\delta}}{\delta^{\circ}}\Bigr )^{2}\zeta_{N}
\ll \tau_{0}
\qquad
(N=1,\ldots,N_{1}). \]
This inequality implies that no viscoelastic response is observed an a test (even if $\zeta_{N}$
is of the order of unity) under the condition
\begin{equation}
\frac{\tilde{\kappa}}{\mu \tau_{0}} \ll 1.
\end{equation}
Based on Eq. (76), one can speculate that the absence of time-dependent response 
in experiments on polybutadiene rubber at room temperature (Aksel and H\"{u}bner, 1996) 
and polyacrylate rubber at elevated temperatures (Clarke et al., 2000)
is associated with slow rates of entropy production in these polymers
[small $\tilde{\kappa}$ is Eq. (48)],
whereas noticeable relaxation of stresses observed in natural rubber at $T=80$ $^{\circ}$C 
(Clarke et al., 2000) may be explained by a substantial increase in the 
dimensionless ratio $\tilde{\kappa}/(\mu \tau_{0})$ with temperature.

\section{Concluding remarks}

Constitutive equations have been derived for the time-dependent
response of rubbery polymers at finite strains.
A particle-reinforced elastomer is modeled as a network of long chains 
bridged by permanent junctions (chemical crosslinks, entanglements and
filler particles).
In agreement with the tube concept, a strand between two neighboring junctions 
is supposed to be confined to a tube which describes constrains on its micro-motion
imposed by surrounding macromolecules.
Unlike the conventional approach that postulates a constant cross-section of the tube, 
its radius is assumed to change substantially with the longitudinal coordinate.
The wavyness of the tube's surface is associated with local heterogeneities of the
polymer (both in space and time) which impose additional constrains on the motion 
of macromolecules.
This implies that a strand may be treated as a sequence of threads (RHMs) whose
thermal motion is unconstrained (in the sense that they go through all possible
configurations during the characteristic time of experiment) and
highly constrained segments with a negligible entropy (which correspond to 
the bottle-neck points of the tube).
Thermal fluctuations result in reconstruction of the tube surface, 
which is reflected by freezing of some RHMs and activation of frozen segments.
The viscoelastic behavior of a rubbery polymer is associated with thermally agitated
changes in the concentrations of RHMs.

The stress--strain relations at finite strains and the kinetic equations for 
the numbers of RHMs in strands are developed by using the laws of thermodynamics.
At small strains, these relations are reduced to the integral 
constitutive equation in linear viscoelasticity (1) with the relaxation kernel 
in the form of a truncated Prony series (2).
Unlike the phenomenological approach which provides no information
about the relaxation spectrum, a novel scaling law (67) is derived 
for the characteristic times of relaxation.
5 adjustable parameters of the model are found by fitting experimental
data in tensile dynamic tests for a carbon black filled natural rubber vulcanizate at various
temperatures, amplitudes and frequencies of oscillations.
Fair agreement is demonstrated between observations and results of numerical simulation.
It is shown that the experimental constants are altered by temperature and amplitude of
straining in a physically plausible way.

\newpage
\section*{References}

\hspace*{6 mm}
Aksel, N., H\"{u}bner, Ch., 1996.
The influence of dewetting in filled elastomers on the changes of
their mechanical properties.
{\em Arch. Appl. Mech.} {\bf 66}, 231--241.

Bergstr\"{o}m, J.S., Boyce, M.C., 1998.
Constitutive modelling of the large strain time-dependent behavior
of elastomers.
{\em J. Mech. Phys. Solids} {\bf 46}, 931--954.

Bergstr\"{o}m, J.S., Boyce, M.C., 2001.
Deformation of elastomeric networks: relation between molecular level deformation 
and classical statistical mechanics models of rubber elasticity.
{\em Macromolecules} {\bf 34}, 614--626.

Clarke, S.M., Elias, F., Terentjev, E.M., 2000.
Ageing of natural rubber under stress.
{\em Eur. Phys. J. E} {\bf 2}, 335--341.

Coleman, B.D., Gurtin, M.E., 1967.
Thermodynamics with internal state variables.
{\em J. Chem. Phys.} {\bf 47}, 597--613.

Doi, M., Edwards, S.F., 1986.
{\em The Theory of Polymer Dynamics.}
Clarendon Press, Oxford.

Drozdov, A.D., 1996.
{\em Finite Elasticity and Viscoelasticity.}
World Scientific, Singapore.

Drozdov, A.D., 1997.
A constitutive model for nonlinear viscoelastic media.
{\em Int. J. Solids Structures} {\bf 34}, 2685-2707.

Dyre, J.C., 1995.
Energy master equation: a low-temperature approximation to B\"{a}ssler's
random-walk model.
{\em Phys. Rev. B} {\bf 51}, 12276--12294.

Ernst, L.J., Septanika, E.G., 1999.
A non-Gaussian network alteration model.
In Dorfmann, A., Muhr, A. (Eds.) 
{\em Constitutive Models for Rubber.}
Balkema, Rotterdam, pp. 169--180.

Everaers, R., 1998.
Constrained fluctuation theories of rubber elasticity:
general results and an exactly solvable model.
{\em Eur. Phys. J. B} {\bf 4}, 341--350.

Filali, M., Michel, E., Mora, S., Molino, F., Porte, G., 2001.
Stress relaxation in model transient networks: percolation and rearrangement
of the crosslinks.
{\em Colloid Surf. A} {\bf 183--185}, 203--212.

Flory, J.D., 1980.
{\em Viscoelastic Properties of Polymers.}
Wiley, New York.

Gonzonas, G.A., 1993.
A uniaxial nonlinear viscoelastic constitutive model with damage for M30 gun propellant.
{\em Mech. Mater.} {\bf 15}, 323--335.

Govindjee, S., Simo, J.C., 1992.
Mullins' effect and the strain amplitude dependence of the
storage modulus.
{\em Int. J. Solids Structures} {\bf 29}, 1737--1751.

Green, M.S., Tobolsky, A.V., 1946.
A new approach to the theory of relaxing polymeric media.
{\em J. Chem. Phys.} {\bf 14}, 80--92.

Ha, K., Schapery, R.A., 1998.
A three-dimensional viscoelastic constitutive model for particulate composites
with growing damage and its experimental validation.
{\em Int. J. Solids Structures} {\bf 35}, 3497--3517.

Haupt, P., Sedlan, K., 2001.
Viscoplasticity of elastomeric materials: experimental facts and constitutive modelling.
{\em Arch. Appl. Mech.} {\bf 71}, 89--109.

Hausler, K., Sayir, M.B., 1995.
Nonlinear viscoelastic response of carbon black reinforced rubber
derived from moderately large deformations in torsion.
{\em J. Mech. Phys. Solids} {\bf 43}, 295--318.

Holzapfel, G., Simo, J., 1996.
A new viscoelastic constitutive model for continuous media
at finite thermomechanical changes.
{\em Int. J. Solids Structures} {\bf 33}, 3019--3034.

Johnson, A.R., Stacer, R.G., 1993.
Rubber viscoelasticity using the physically constrained system's stretches
as internal variables.
{\em Rubber Chem. Technol.} {\bf 66}, 567--577.

Johnson, A.R., Quigley, J., Freese, C.E., 1995.
A viscohyperelastic finite element model for rubber.
{\em Comput. Methods Appl. Mech. Engng.} {\bf 127}, 163--180.

Kar, K.K. Bhowmick, A.K., 1997.
Hysteresis loss in filled rubber vulcanizates and its relationship
with heat generation.
{\em J. Appl. Polym. Sci.} {\bf 64}, 1541--1555.

Leblanc, J.L., Cartault, M., 2001.
Advanced torsional dynamic methods to study the morphology of uncured filled rubber 
compounds.
{\em J. Appl. Polym. Sci.} {\bf 80}, 2093--2104.

Le Tallec, P., Rahier, C., Kaiss, A., 1993.
Three-dimensional incompressible viscoelasticity in large strains:
formulation and numerical approximation.
{\em Comp. Meths. Appl. Mech. Engng.} {\bf 109}, 233--258.

Liang, J.Z., Li, R.K.Y., Tjong, S.C., 1999.
Effects of glass bead content and surface treatment on viscoelasticity
of filled polypropylene/elastomer hybrid composites.
{\em Polym. Int.} {\bf 48}, 1068--1072.

Lion, A., 1996.
A constitutive model for carbon black filled rubber:
experimental investigations and mathematical representation.
{\em Continuum Mech. Thermodyn.} {\bf 8}, 153--169.

Lion, A., 1997.
On the large deformation behavior of reinforced rubber at different
temperatures.
{\em J. Mech. Phys. Solids} {\bf 45}, 1805--1834.

Lion, A., 1998.
Thixotropic behaviour of rubber under dynamic loading histories:
experiments and theory.
{\em J. Mech. Phys. Solids} {\bf 46}, 895--930.

Lodge, A.S., 1968.
Constitutive equations from molecular network theories
for polymer solutions.
{\em Rheol. Acta} {\bf 7}, 379--392.

Miehe, C., Keck, J., 2000.
Superimposed finite elastic-viscoelastic-plastoelastic
stress response with damage in filled rubbery polymers.
Experiments, modelling and algorithmic implementation.
{\em J. Mech. Phys. Solids} {\bf 48}, 323--365.

Ozupek, S., Becker, E.B., 1992.
Constitutive modeling of high-elongation solid propellant.
{\em J. Engng. Mater. Technol.} {\ bf 114}, 111--115.

Palmas, P., Le Campion, L., Bourgeoisat, C., Martel, L., 2001.
Curing and thermal ageing of elastomers as studied by $^{1}$H broadband
and $^{13}$C high-resolution solid-state NMR.
{\em Polymer} {\bf 42}, 7675--7683.

Reese, S., Govinjee, S., 1998.
Theoretical and numerical aspects in the thermo-viscoelastic material
behaviour of rubber-like polymers.
{\em Mech. Time-Dependent Mater.} {\bf 1}, 357--396.

Septanika, E.G., Ernst, L.J., 1998a.
Application of the network alteration theory for modeling the
time-dependent constitutive behaviour of rubbers.
1. General theory.
{\em Mech. Mater.} {\bf 30}, 253--263.

Septanika, E.G., Ernst, L.J., 1998b.
Application of the network alteration theory for modeling the
time-dependent constitutive behaviour of rubbers.
2. Further evaluation of the general theory and experimental
verification.
{\em Mech. Mater.} {\bf 30}, 265--273.

Spathis, G., 1997.
Non-linear constitutive equations for viscoelastic behaviour of
elastomers at large deformations.
{\em Polym. Gels Networks} {\bf 5}, 55--68.

Sperling, L.H., 1996.
{\em Introduction to Physical Polymer Science.}
Wiley, New York.

Tanaka, F., Edwards, S.F., 1992.
Viscoelastic properties of physically cross-linked networks.
Transient network theory.
{\em Macromolecules} {\bf 25}, 1516--1523.

Treloar, L.R.G., 1975.
{\em The Physics of Rubber Elasticity.}
Clarendon Press, Oxford.

Wu, J.-D., Liechti, K.M., 2000.
Multiaxial and time dependent behavior of a filled rubber.
{\em Mech. Time-Dependent Mater.} {\bf 4}, 293--331.

Yamamoto, M., 1956.
The visco-elastic properties of network structure.
1. General formalism.
{\em J. Phys. Soc. Japan} {\bf 11}, 413--421.

Zdunek, A.B., 1993.
Theory and computation of the steady state harmonic response of
viscoelastic rubber parts.
{\em Comput. Methods Appl. Mech. Engng.} {\bf 105}, 63--92.
\newpage

\begin{center}
Table 1: Adjustable parameters $\tilde{N}_{i}$ and $\Sigma_{i}$ at various
temperatures
\vspace*{6 mm}

\begin{tabular}{@{} c c r c r @{}}
\hline
$T$ K & $\tilde{N}_{0}$ & $\tilde{N}_{1}$ & $\Sigma_{0}$ & $\Sigma_{1}$ \cr
\hline
253.0 & 127.77 & 702.07  & 40.52 & $-16.46$\\
296.0 & 141.70 & 30.66   & 31.00 & 0.00\\
333.0 & 140.00 & 315.69  & 43.58 & 43.58\\
373.0 & 115.22 & 3139.68 & 25.34 & 600.29\\
\hline
\end{tabular}
\end{center}

\begin{center}
Table 2: Adjustable parameters $\tilde{N}$ and $\Sigma$ for oscillation tests 
with the amplitude $\Delta=0.006$
\vspace*{6 mm}

%
\end{center}

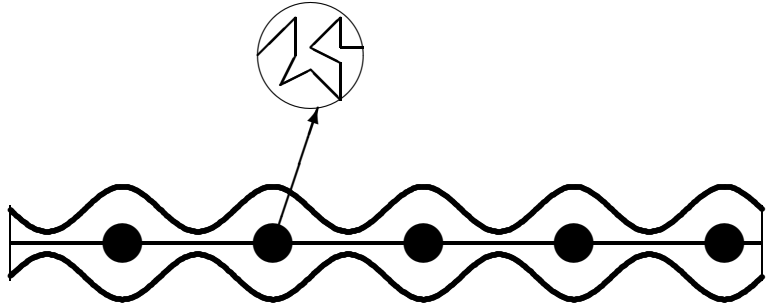
\caption{A piece of a strand confined to a wavy tube.
Filled circles: regions of high molecular mobility; intervals between them: 
regions of suppressed mobility}
\end{figure}

\setlength{\unitlength}{0.8 mm}
\begin{figure}[t]
\begin{center}
%
\end{center}
\vspace*{10 mm}

\caption{The storage modulus $E^{\prime}$ MPa versus the frequency of
oscillations $f$ Hz in a dynamic test with the strain amplitude $\Delta$
at $T=253$ K.
Circles: experimental data (Lion, 1998).
Solid lines: results of numerical simulation.
Curve 1: $\Delta=0.006$;
curve 2: $\Delta=0.011$;
curve 3: $\Delta=0.028$;
curve 4: $\Delta=0.056$}
\end{figure}

\begin{figure}[t]
\begin{center}
%
\end{center}
\vspace*{10 mm}

\caption{The storage modulus $E^{\prime}$ MPa versus the frequency of
oscillations $f$ Hz in a dynamic test with the strain amplitude $\Delta$
at $T=296$ K.
Circles: experimental data (Lion, 1998).
Solid lines: results of numerical simulation.
Curve 1: $\Delta=0.006$;
curve 2: $\Delta=0.011$;
curve 3: $\Delta=0.028$;
curve 4: $\Delta=0.056$}
\end{figure}

\begin{figure}[t]
\begin{center}
%
\end{center}
\vspace*{10 mm}

\caption{The storage modulus $E^{\prime}$ MPa versus the frequency of
oscillations $f$ Hz in a dynamic test with the strain amplitude $\Delta$
at $T=333$ K.
Circles: experimental data (Lion, 1998).
Solid lines: results of numerical simulation.
Curve 1: $\Delta=0.006$;
curve 2: $\Delta=0.011$;
curve 3: $\Delta=0.028$;
curve 4: $\Delta=0.056$}
\end{figure}

\begin{figure}[t]
\begin{center}
%
\end{center}
\vspace*{10 mm}

\caption{The average number of RHMs in a strand $\tilde{N}$ 
versus the amplitude of oscillations $\Delta$ at $T=253$ K.
Cirles: treatment of observations (Lion, 1998).
Solid line: approximation of the experimental data by Eq. (69)}
\end{figure}

\begin{figure}[t]
\begin{center}
%
\end{center}
\vspace*{10 mm}

\caption{The average number of RHMs in a strand $\tilde{N}$ 
versus the amplitude of oscillations $\Delta$ at $T=296$ K.
Cirles: treatment of observations (Lion, 1998).
Solid line: approximation of the experimental data by Eq. (69)}
\end{figure}

\begin{figure}[t]
\begin{center}
%
\end{center}
\vspace*{10 mm}

\caption{The average number of RHMs in a strand $\tilde{N}$ 
versus the amplitude of oscillations $\Delta$ at $T=333$ K.
Cirles: treatment of observations (Lion, 1998).
Solid line: approximation of the experimental data by Eq. (69)}
\end{figure}

\begin{figure}[t]
\begin{center}
%
\end{center}
\vspace*{10 mm}

\caption{The standard deviation of the number of RHMs in a strand
$\Sigma$ versus the amplitude of oscillations $\Delta$ at $T=253$ K.
Cirles: treatment of observations (Lion, 1998).
Solid line: approximation of the experimental data by Eq. (70)}
\end{figure}

\begin{figure}[t]
\begin{center}
%
\end{center}
\vspace*{10 mm}

\caption{The standard deviation of the number of RHMs in a strand
$\Sigma$ versus the amplitude of oscillations $\Delta$ at $T=296$ K.
Cirles: treatment of observations (Lion, 1998).
Solid line: approximation of the experimental data by Eq. (70)}
\end{figure}

\begin{figure}[t]
\begin{center}
%
\end{center}
\vspace*{10 mm}

\caption{The standard deviation of the number of RHMs in a strand
$\Sigma$ versus the amplitude of oscillations $\Delta$ at $T=333$ K.
Cirles: treatment of observations (Lion, 1998).
Solid line: approximation of the experimental data by Eq. (70)}
\end{figure}

\begin{figure}[t]
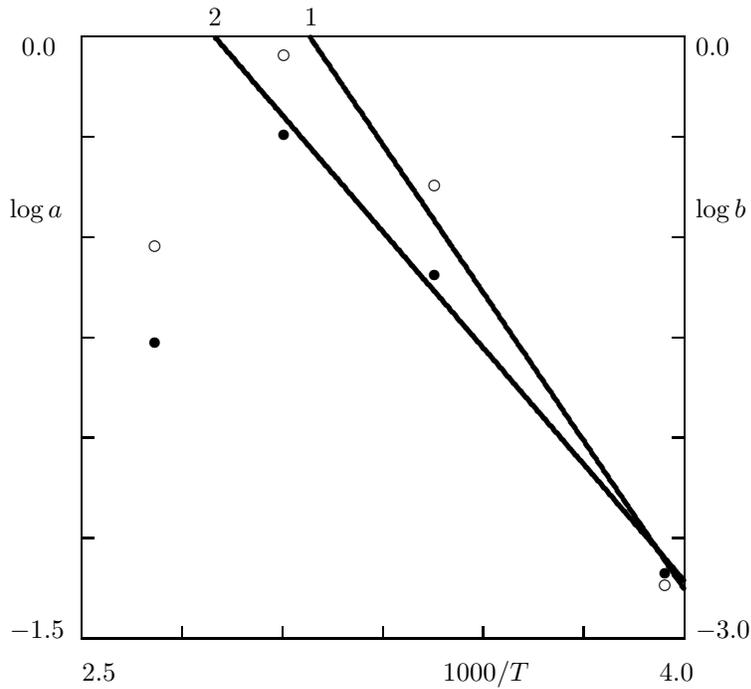

\begin{center}
%
\end{center}
\vspace*{10 mm}

\caption{The parameters $a$ s$^{-1}$ (unfilled circles)
and $b$ s$^{-1}$ (filled circles) versus temperature $T$ K.
Cirles: treatment of observations (Lion, 1998).
Solid lines: approximation of the experimental data by Eq. (71).
Curve 1: $a_{0}=4.5289$, $a_{1}=1.4759\cdot 10^{3}$;
curve 2: $b_{0}=6.5638$, $b_{1}=2.3182\cdot 10^{3}$}
\end{figure}

\end{document}